\documentstyle[12pt,aaspp4]{article}

\newcommand{\kms}{$\,\mbox{km}\,\mbox{s}^{-1}$}
\newcommand{\mo}{\mbox{M}$_\odot$}
\newcommand{\nullpos}{$\alpha_{1950}=9^h59^m57.\!^s3$, $\delta_{1950}=68^\circ57'13.\!''8$}

\slugcomment{accepted, ApJ Letters}

\lefthead{Walter \& Heithausen}
\righthead{Molecular Gas near NGC\,3077}

\begin{document}

\title {The Discovery of a Molecular Complex in the Tidal Arms near NGC\,3077}

  \author {Fabian Walter and Andreas Heithausen }
  \affil {Radioastronomisches Institut der Universit\"at Bonn, Auf
dem H\"ugel 71, 53121 Bonn, Germany;\\
walter@astro.uni--bonn.de; heith@astro.uni--bonn.de}

%
%

\begin{abstract}

We present the discovery of a giant molecular complex
(${r\approx350}$\,pc, ${M_{\rm vir}\! \approx \! 10^7}$\,\mo) in the
tidal arms south--east of NGC\,3077, a member of the M\,81
triplet. The complex is clearly detected in the $^{12}$CO (${J=1\! \to
\!0}$) transition at five independent positions. The position relative
to NGC\,3077, the systemic velocity ($v_{\rm hel}\approx$14\kms) and
the cloud averaged line width ($\Delta v\approx$15\kms) indicate that
the object is not related to Galactic cirrus but is extragalactic. The
tidal \ion{H}{1} arm where the molecular complex is located has an
total \ion{H}{1} mass of $M_{\rm HI}\approx 3 \times 10^8$\,\mo. This
tidal material was presumably stripped off the outer parts of
NGC\,3077 during the closest encounter with M\,81, about $3 \times
10^8$ years ago.  After the complex detected along a torn-out spiral
arm of M\,81 by Brouillet et al., it is the second of its kind
reported so far. Based on published optical observations, we have no
evidence for on--going star formation in the newly detected molecular
complex.  Since the system has all the ingredients to form stars in
the future, we speculate that it might eventually resemble the young
dwarf galaxies in the M\,81 group.

\end{abstract}

%
%


%
%

\newpage

\section{Introduction\label{introduction}}

Since about a decade it has been established that dwarf galaxies are
the most numerous type of galaxies in the universe (e.g.\ Mateo
1998)\markcite{mateo98}.  Zwicky (1956)\markcite{zwicky56} was among
the first to propose that at least some dwarf galaxies might have been
created from tidal debris strewn about in intergalactic space by
gravitational encounters between massive spiral galaxies. The
relatively nearby M\,81 triplet, for example, is one of the most
stunning examples of how gravitational interactions can redistribute
the neutral atomic gas, originally belonging to individual galaxies,
over a huge volume in the form of bridges and tidal tails (van der
Hulst 1979\markcite{vanderhulst79}, Yun et al.\
1993\markcite{yun:etal93}, Yun, Ho \& Lo\
1994)\markcite{yun:etal94}. The newly born galaxies which may form in
these types of interactions are variously referred to as
`protogalaxies', i.e., those who have not yet started to form stars,
`tidal dwarfs' or `intergalactic molecular complexes'. Recently, they
have also been coined `Phoenix galaxies' (Hernquist
1992\markcite{hernquist92}) since they are presumably born
phoenix--like out of the `ashes' of galaxy--galaxy collisions.

The recent discovery of star forming regions in tidal tails of
interacting galaxies (Mirabel, Lutz \& Maza\
1991\markcite{mirabel:etal91}, Mirabel, Dottori \& Lutz
1992\markcite{mirabel:etal92}; Duc et al.\ 1997\markcite{duc:etal97};
Duc \& Mirabel 1998\markcite{duc:mirabel98}) provides support for the
concept of Phoenix--galaxies. This is corroborated by numerical
simulations which show that tails produced by tidal interactions are
subject to fragmentation (Barnes \& Hernquist
1992\markcite{barnes:hernquist96}; Elmegreen, Kaufman \& Thomasson
1993\markcite{elmegreen:etal93}) which may lead to star formation.

Despite the fact that recent star formation is taking place in tidal
dwarfs, searches for molecular gas in these objects have been rather
unsuccessful thus far. This is surprising as, according to our current
understanding of the star formation process, neutral gas needs to
get dense enough, become self gravitating, and turn largely from
the atomic to a denser molecular phase before star formation can
proceed. Because molecular hydrogen, the main constituent of a
molecular cloud, is difficult to detect at the temperatures and
densities found in molecular clouds, emission from carbon monoxide
(CO) is commonly used as a tracer.

The only example to date of a molecular complex associated with tidal
debris is the one detected by Brouillet et al.\ (1992) within the
M\,81 triplet, close to, but outside the optical image of
M\,81. Although it is presumably located within an \ion{H}{1} arm torn
out of M\,81, no optical emission or star formation appears to be
associated with it (Henkel et al.\
1993\markcite{henkel:etal93}). Since this discovery, a lot of
observational effort has been undertaken to detect similar molecular
complexes in tidal arms of interacting groups of galaxies -- but
without success (e.g., Brouillet, Henkel \& Baudry
1992\markcite{brouillet:etal92}, Smith \& Higdon
1994\markcite{smith:higdon94}). In this paper, we present the
detection of the second molecular complex of this kind.

%
%

\section{Observations\label{observations}}

We searched for molecular gas in the tidal arms near NGC\,3077 which
show their presence in observations of the 21\,cm line of neutral
hydrogen (\ion{H}{1}). NGC\,3077 is a member of the nearby
(D=3.2\,Mpc) interacting M\,81 galaxy triplet (see Fig.\ 1, left).
Our CO observations were guided by archival \ion{H}{1} data of
NGC\,3077 (Fig.~1, right, greyscale) obtained with the NRAO Very Large
Array (VLA)\footnote{The National Radio Astronomy Observatory (NRAO)
is operated by Associated Universities, Inc., under cooperative
agreement with the National Science Foundation.} with an improved
angular resolution (10$''$) of a factor of 6 as compared to published
\ion{H}{1} observations (van der Hulst 1979\markcite{vanderhulst79};
Yun et al.\ 1993\markcite{yun:etal93}).  The increased resolution
makes it easier to identify regions of high \ion{H}{1} column density
and hence provides better clues as to where to search for molecular
clouds.

The CO observations were carried out in December 1998 with the IRAM
30\, m radio telescope. We observed the $^{12}$CO (${J=1\to0}$) and
(${J=2\! \to \! 1}$) transitions at 115\, GHz and 230\, GHz
simultaneously using two dual channel receivers with a wobbling
subreflector. The wobbler throw was $\pm4'$ in azimuth.  The angular
resolution of the telescope is 11$''$ at 230\, GHz and 22$''$ at 115\,
GHz. In total, 13 independent positions were observed within 6 hours
observing time on a $20''$ grid, which is approximately the size of
the telescope beam at 115\,GHz. We used the 1 MHz filter spectrometer
and the autocorrelator spectrometer simultaneously. Spectra for
individual positions were finally degraded to a velocity resolution of
2.6\,\kms\ at both frequencies. The different receiver setups were
combined to one single spectrum per position and transition. Only
linear baselines were removed from the spectra. The final rms ($T_{\rm
mb}$) values for individual positions are listed in Tab.~1.

\begin{deluxetable}{cccccc}
\tablewidth{0pc}
\tablecaption{Gaussian components of the CO ($J=1\!\to\! 0$) spectra.}  
\tablehead{
\colhead{$\Delta\alpha$\tablenotemark{a}} &
\colhead{$\Delta\delta$\tablenotemark{a}} & 
\colhead{$T_{\rm mb}$} &
\colhead{rms} & 
\colhead{$v_{\rm hel}$} & 
\colhead{$\Delta v$}  \nl
\colhead{$('')$} &
\colhead{$('')$} & 
\colhead{(mK)} &
\colhead{(mK)} & 
\colhead{(\kms)}&  
\colhead{(\kms)}
}
\startdata
    0& 60 & --  & 14.0 & $-$ & $-$ \nl
    0& 40 & --  & 15.9 & $-$ & $-$ \nl
 --20& 20 & --  &  9.6 & $-$ & $-$ \nl
    0& 20 & 72.8& 15.1 & $13.9\pm0.9$ & $15.3\pm2.1$ \nl
   20& 20 & 47.8& 13.1 & $14.4\pm1.1$ & $12.5\pm3.0$ \nl
 --40& 0  & 29.7& 14.8 & $ 9.2\pm2.8$ & $21.7\pm5.5$ \nl
 --20& 0  & 43.9& 10.8 & $11.9\pm1.4$ & $23.9\pm3.3$ \nl
    0& 0  & 59.4&  8.1 & $15.0\pm0.7$ & $15.1\pm1.6$ \nl
   20& 0  & --  & 13.5 & $-$ & $-$ \nl
   40& 0  & --  & 13.4 & $-$ & $-$ \nl
 --20&--20& --  & 10.9 & $-$ & $-$ \nl
    0&--20& 46.0& 14.4 & $14.2\pm1.0$ & $8.5\pm1.9$ \nl
    0&--60&  -- & 13.9 & $-$ & $-$ \nl
\enddata
\tablenotetext{a}{Positions are offsets relative to \nullpos.}
\label{gauss}
\end{deluxetable}

\begin{figure}
\plotone{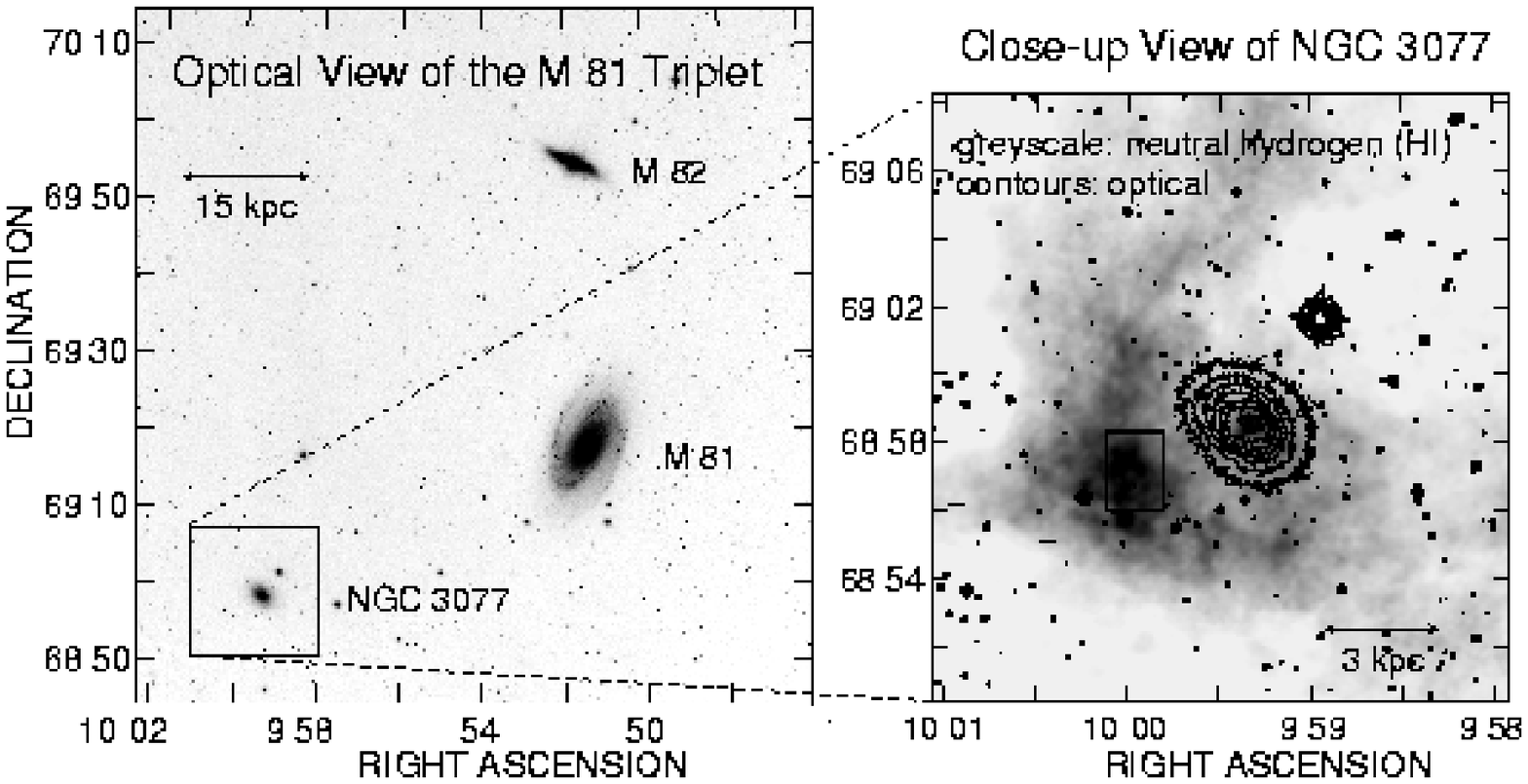}
\caption{ {\em Left:} Optical view of the M\,81--M\,82--NGC\,3077
triplet (taken from the Digitized Sky Survey). A linear scale
(adopting a distance to the triplet of D=3.2\,Mpc) is indicated in the
upper left. The box marks the region around NGC\,3077 which is blown
up on the right. {\em Right:}~Close--up view of the region around
NGC\,3077. The greyscale represents the distribution of the neutral
hydrogen (\ion{H}{1}) as observed with the VLA.  The contours
represent the optical image (shown in greyscale on the left). The box
indicates the region where molecular gas was discovered (see Fig.~2
for a close--up view of that region). The linear scale is indicated in
the lower right. Coordinates are given in the B1950.0 epoch.
\label{overview}}
\end{figure}

%
%

\section{Results}

Our CO spectra are presented in Figs.\ 2 and 3. Fig.\ 2 gives an
overview of our limited mapping result towards the newly detected
molecular complex.  CO ($J=1\to0$) emission was clearly detected from
the central five positions towards the region with the highest
\ion{H}{1} column densities.  There is evidence of more emission in
neighboring spectra, but at a lower than 3\,$\sigma$ confidence level.
The properties of the CO ($J=1\to0$) lines as derived from a Gaussian
analysis are listed in Tab.~1.  Brouillet et al.\
(1992)\markcite{brouillet:etal92} obtained two spectra close to the
region we mapped. They are ($-2''$, $-73''$) and (240$''$, 16$''$)
offset from our reference position \nullpos.  However, Brouillet et
al.\ did not detect CO emission in their spectra at a 1\,$\sigma$--rms
of about 30\, mK in their 5.2\,\kms\ wide channels.  Due to the lower
signal-to-noise ratio, emission from the CO ($J=2\to1$) was only
unambiguously detected from our (0,0) position, where we integrated
longest, and for the average spectrum of the five positions where we
detected the lower ($J=1\to0$) transition.  For two other positions
where we detected the ($1\to0$) transition there is indication for
($2\to1$) emission also, however with less than 3$\sigma$.

Fig.\ 3 presents a comparison between the spectra of neutral hydrogen
(\ion{H}{1}) and the two lowest rotational CO transitions towards our
reference position. The velocities of the molecular gas in the newly
detected molecular complex are in excellent agreement with the atomic
component. As is also found in other galaxies, the \ion{H}{1} line is
broader than the CO line, indicating that the molecular material is
clumped and present over a smaller range along the line of sight
than the \ion{H}{1}.

The ratio of the two integrated CO lines is (${W_{2\to1}/W_{1\to0})
=0.75\pm0.12}$ for the average of the central five positions; because
the source is clearly extended (s. Fig. 2) no correction for the
different beam areas for the two transitions was made.  This ratio is
similar to that found in our Galaxy for quiescent cool (10--20\,K)
molecular clouds (Falgarone et al.\ 1998)\markcite{falgarone:etal98}.

\begin{figure}
\epsscale{0.5}
\plotone{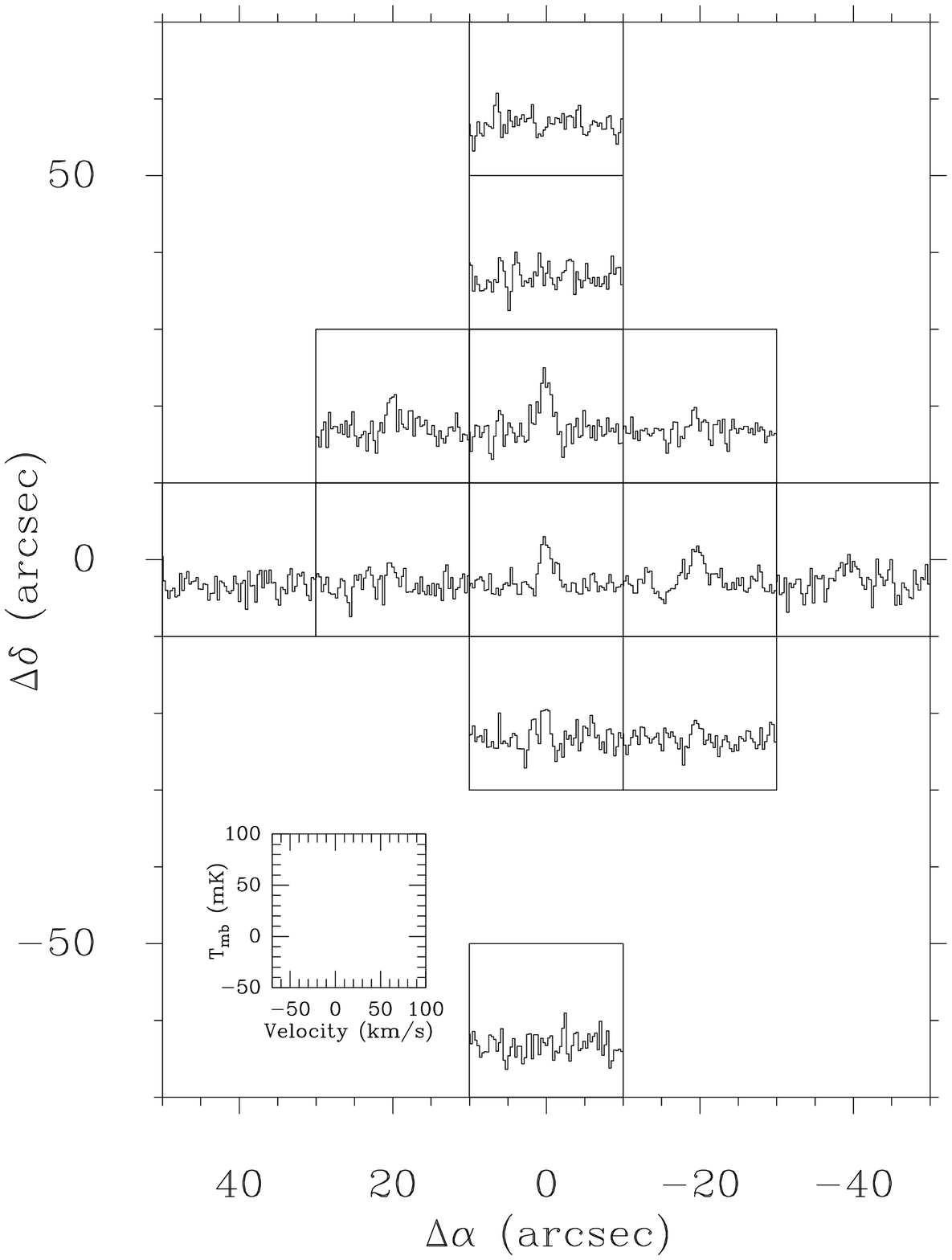}
\label{spectramap}
\caption{$^{12}$CO (${J=1\!\to\!0}$) spectra obtained towards the
newly detected molecular complex.  Positions are offsets relative to
\nullpos. The center of each square corresponds to the location where
each displayed spectrum has been obtained. Note that the spacing
between two individual spectra (20$''$) is approximately the size of
the IRAM 30\,m telescope at 115 GHz (22$''$).  The small inserted box
indicates the velocity and temperature scale for each spectrum.}
\end{figure}

\begin{figure}
\epsscale{0.5}
\plotone{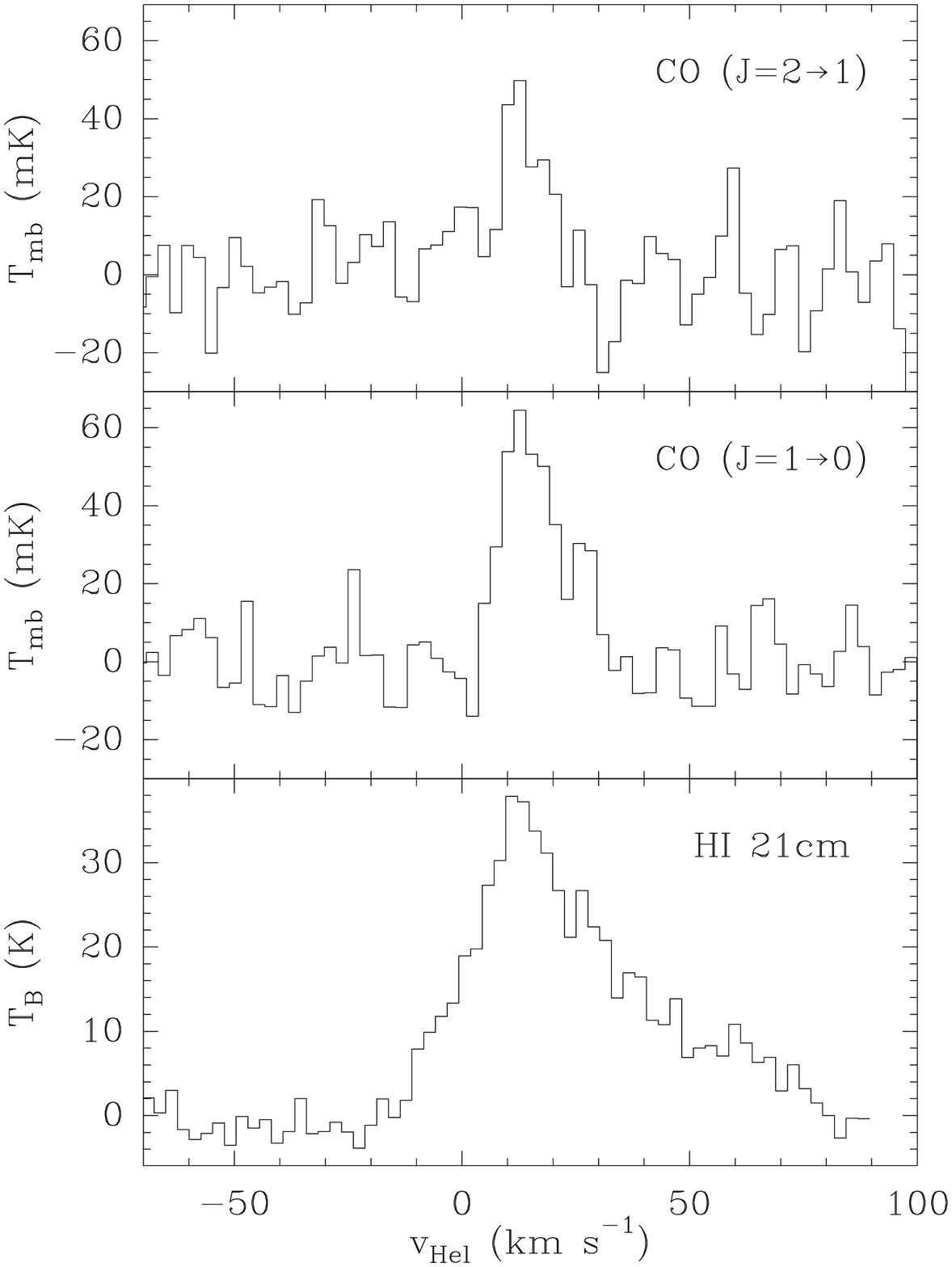}
\label{spectramap}
\caption{$^{12}$CO (${J=1\!\to\!0}$) spectra obtained towards the
newly detected molecular complex.  Positions are offsets relative to
\nullpos. The center of each square corresponds to the location where
each displayed spectrum has been obtained. Note that the spacing
between two individual spectra (20$''$) is approximately the size of
the IRAM 30\,m telescope at 115 GHz (22$''$).  The small inserted box
indicates the velocity and temperature scale for each spectrum.}
\end{figure}

%
%

\section{Discussion\label{discussion}}

\subsection{The properties of the molecular complex}

Because the CO detection coincides with the peak in the \ion{H}{1}
distribution of the tidal gas near NGC\,3077, and strengthened by the
fact that the velocities of the \ion{H}{1} and CO are in excellent
agreement, it is natural to assume that the newly detected molecular
gas is located within the M\,81--M\,82--NGC\,3077 triplet. However, it
is important to further rule out a possible Galactic origin.

The molecular gas is located towards a direction where the Galaxy is
rich in infrared cirrus (de Vries, Heithausen \& Thaddeus\
1987\markcite{devries:etal87}).  A significant part of these cirrus
clouds is molecular.  However, these clouds are distinct from the
cloud we detected for two reasons.  First, molecular cirrus clouds in
this region of the sky have more negative velocities (de Vries et al.\
1987\markcite{devries:etal87}), the closest cirrus cloud is at $v_{\rm
hel}=-4.5$\,\kms\ thus 18.5\,\kms\ off our CO detection.  Secondly,
the line width of our molecular cloud ($15\pm1$\kms) is significantly
broader than those of the cirrus clouds ($\Delta v\le4.4$\,\kms, cf.\
Heithausen 1996\markcite{heithausen96}) and is more similar to that of
giant molecular complexes in our Galaxy.  We thus conclude that our
object is not connected to foreground cirrus clouds but extragalactic
in nature.

We therefore assume that the molecular complex is indeed associated
with the tidal debris of NGC\,3077.  We adopt a distance of 3.2\, Mpc
(cf.\ Henkel et al.\ 1993\markcite{henkel:etal93}). As shown in Fig.\
1, the region where we detected CO is clearly outside the optical
extent of NGC\,3077. The radius ($R_{25}$) of NGC\,3077 is
$2.\!'7\times2.\!'2$ or 2.5\, kpc $\times$ 2.1\, kpc (de Vaucouleurs
et al.\ 1991\markcite{devaucouleurs:etal91}). Molecular gas of
NGC\,3077 itself is concentrated to its nucleus with a half width at
half maximum (HWHM) of about 0.16\,kpc (Becker, Schilke, \& Henkel
1989)\markcite{becker:etal89}, atomic gas of NGC\,3077 extends further
out (HWHM$\,\approx\,$0.9\,kpc).  The projected distance of the newly
detected molecular complex to the center of NGC\,3077 is $3.\!'9$
corresponding to 3.7\, kpc at the distance of NGC\,3077.

If we assume our limited mapping to be complete, molecular gas in the
complex covers an area of 0.4\, kpc$^2$, which corresponds to an
effective radius of about 350\, pc.  This is about twice the value
found by Brouillet et al.\ (1992)\markcite{brouillet:etal92} for their
intergalactic complex near M\,81. We note, however, that due to the
low line--intensities observed, we expect beam dilution to play a
major role. This means that the molecular gas in the complex is likely
clumped.

For comparison purpose we calculate a virial mass of the molecular
complex of $M_{\rm vir}\approx 1.7\times10^7$\,\mo, assuming a
constant density throughout the molecular gas. This value is similar
to the object found by Brouillet et al.\
(1992)\markcite{brouillet:etal92} and also similar to the molecular
complex near the nucleus of NGC\,3077 (Becker et al.\
1989\markcite{becker:etal89}). Adopting a Galactic $X$--factor, the
ratio of H$_2$ column density to integrated CO intensity, ${X_{\rm
G}=(1.56\pm0.05)\times10^{20}}$\, cm$^{-2}$\,(K\,\kms)$^{-1}$ (Hunter
et al.\ 1997\markcite{hunter:etal97}), we derive a molecular mass
which is a factor of 20 lower than that based on the assumption of
virialisation. This rather large discrepancy in the mass estimates
could have two reasons: i) the assumption of virialisation does not
hold for this complex; ii) the true $X-$factor in the molecular
complex is comparable to values found in low metallicity galaxies
(Dettmar \& Heithausen 1989\markcite{dettmar:heithausen89}) and in the
Brouillet et al.\ complex. Observations of, e.g., rarer CO isotopomers
are needed to investigate this issue further.

\subsection{The formation of the molecular complex}

The metallicity of interstellar gas can provide important clues as to
the origin of that gas.  From the mere fact that we see CO we can
already conclude that the material in the molecular complex near
NGC\,3077 is not primordial, but already pre--processed.

Yun et al.\ (1993)\markcite{yun:etal93} performed detailed numerical
simulations to explain the impressive tidal \ion{H}{1}--structure in
the M\,81 triplet.  These simulations suggest that the huge \ion{H}{1}
complex east of NGC\,3077 (total \ion{H}{1} mass: $M_{\rm
HI}=3\times10^8$\mo) where the molecular complex is situated was
formed out of material stripped off the outer parts of NGC\,3077
during the closest encounter with the most massive galaxy of the
triplet, M\,81, about $3\times 10^8$ years ago.  Optical studies show
that NGC\,3077's metallicity is around solar (Martin
1997)\markcite{martin97}.  This means that even the gas in the
outskirts of NGC\,3077 was already chemically enriched before the
encounter with M\,81 which explains why heavy elements are present.

\subsection{Signs of star formation?}

At this point it is interesting to investigate whether stars have
already formed in the newly detected molecular complex near NGC\,3077
or not. In the case of the intergalactic complex detected by Brouillet
et al.\ (1992)\markcite{brouillet:etal92}, deep optical follow-up
studies showed that no stellar counterpart was
visible\markcite{henkel:etal93}.  In our case, an optical study of
this region using the Russian 6\,m telescope has been undertaken by
Karachentsev, Karachentseva \& B\"orngen\
(1985a)\markcite{karachentsev:etal85a} who obtained deep images of an
irregular dwarf object south--east of NGC\,3077 which they named the
`Garland'. As the name indicates, the Garland is a chain of faint
optical knots which occupy an area of $6'\times 4'$ on the sky. Some
diffuse emission is also present in and near the region we mapped.
Karachentsev, Karachenseva \& B\"orngen\
(1985b)\markcite{karachentsev:etal85b} also performed optical
spectroscopy of the four brightest knots in the Garland and determined
a velocity of $(55\pm20)$\kms\ for the knot which is nearest to the
center of our molecular complex ($30''$ to the southwest).  The offset
in velocity relative to the complex suggests that the two systems are
probably not associated with each other.  The observations therefore
suggest that star formation did not proceed as yet in the newly
detected molecular complex.  However, even deeper optical follow up
observations are needed to answer this question unambiguously.

%
%

\section{Conclusions\label{conclusions}}

We have established the presence of a giant cold and quiescent
intergalactic molecular complex with no evidence of on--going star
formation.  The newly detected molecular complex is embedded in a
large region containing \ion{H}{1} gas and thus has all the
ingredients to form a dwarf galaxy in the future. This object may
therefore be classified as a `Phoenix' galaxy and might resemble the
young dwarf galaxies in the M\,81 group in the future.  Our discovery
shows that molecular complexes in tidal arms of interacting galaxies
are not necessarily rare objects --- although they are difficult to
find.

\acknowledgments We thank an anonymous referee for valuable comments
which helped to improve this Letter.  FW acknowledges the 'Deutsche
Forschungsgemeinschaft (DFG)' for the award of a stipendium in the
Graduate School "The Magellanic Clouds and other Dwarf Galaxies". We
thank Christian Henkel, Elias Brinks, and Klaas de Boer for fruitful
discussions.

%
%


\end{document}